\documentstyle[12pt]{article}

\newcommand{\lsim}   {\mathrel{\mathop{\kern 0pt \rlap
  {\raise.2ex\hbox{$<$}}}
  \lower.9ex\hbox{\kern-.190em $\sim$}}}
\newcommand{\gsim}   {\mathrel{\mathop{\kern 0pt \rlap
  {\raise.2ex\hbox{$>$}}}
  \lower.9ex\hbox{\kern-.190em $\sim$}}}

\begin{document}
\input epsf \renewcommand{\topfraction}{0.8}
\pagestyle{empty} \vspace*{5mm}
\begin{center}
\Large{\bf  Could dark matter or neutrinos discriminate between
the enantiomers of a chiral molecule?}
\\ \vspace*{2cm}
\large{ Pedro Bargue\~no}  \\
\vspace{0.2cm}
\normalsize
 Instituto de Matem\'aticas y F\'{\i}sica Fundamental, CSIC, Madrid, Spain\\
Departamento de Qu\'{\i}mica F\'{\i}sica, Universidad de Salamanca, Spain\\
\vspace*{0.6cm}
\large{Antonio Dobado and Isabel Gonzalo } \\
\vspace{0.2cm} \normalsize
Departamento de F\'{\i}sica Te\'orica I
and Departamento de
\'Optica \\
 Universidad  Complutense de Madrid, 28040 Madrid, Spain
\vspace*{0.6cm}

 {\bf ABSTRACT} \\

\end{center}

We examine the effect of cold dark matter on the discrimination
between the two enantiomers of a chiral molecule. We estimate the
energy difference between the two enantiomers due to the
interaction between fermionic WIMPs (weak interacting massive
particles) and molecular electrons on the basis that electrons
have opposite helicities in opposite enantiomers. It is found that
this energy difference is completely negligible. Dark matter could
then be discarded as an inductor of chiroselection between
enantiomers and then of biological homochirality. However, the
effect of cosmological neutrinos, revisited with the currently
accepted neutrino density, would
reach, in the most favorable case, an upper bound of the same
order of magnitude as
the energy difference obtained from the well known electroweak
electron-nucleus interaction in some molecules.

\vspace*{5mm}

\noindent
\begin{flushleft} PACS: 33.15.Bh, 13.15.+g, 95.35.+d\\
\end{flushleft}
\newpage
\setcounter{page}{1} \pagestyle{plain}
\textheight 20 true cm

\section{Introduction}

The origin of biological homochirality, that is, the almost
exclusive one-handedness of  chiral molecules in  biological
organisms, is a fundamental problem for which there is not yet a
convincing solution. Several mechanisms have been proposed to
explain chiroselection among the two possible enantiomers of a
chiral molecule (see for example
\cite{Bonner91,Jorissen02,Clinebook} and references therein).
These mechanisms involve chance, $\beta$-radiolysis
\cite{Hegstrom85}, circularly polarized light
\cite{Bailey98,Bailey01,Buscher05}, magnetic fields
\cite{Rikken00,Ruchon06}, and violation of parity in the weak
interaction (see below).

The discovery of an excess of L-amino acids in meteorites \cite{
Engel97,Pizzarello00} has reinforced the idea of an
extraterrestrial origin of biological homochirality
\cite{Bonner91,Bonner92}. In this context, universal mechanisms of
chiroselection such as parity violation in weak interactions would
acquire special interest in spite their tiny effects, without of
course underestimating other mechanisms.

The effect of electroweak interactions between electrons and
nuclei mediated by the Z$^0$, have been extensively studied and
observed in atoms (see the review \cite{Bouchiat97}), and only
predicted in molecules, where an energy difference between the two
enantiomers of chiral molecules has been estimated to be between
$10^{-16}$ and $10^{-21}$ eV
\cite{Letokhov75,Rein80,Zanasi99,Laerdahl00,Soulard06}. In the
laboratory, no conclusive energy difference have been reported in
experimental spectroscopic studies reaching an energy resolution
of about $10^{-15}$ eV \cite{Crass05}.

The above tiny energy difference would require a powerful
mechanism of amplification in order to induce a real
enantioselective effect. Otherwise the small energy difference
would be masked by the natural broadening of the energy levels of
the molecule, thermal fluctuations and environment interactions,
which do not discriminate, in average, between L and D
enantiomers. There is active research on amplification mechanisms
in which a permanent although very small interaction acting always
in the same enantioselective direction, and under appropriate
conditions, could lead to an effective enantioselection. Some
mechanisms are based on nonlinear autocatalytic processes of
polymerization or crystallization along a large period of time
\cite{Kondepudi85,Kondepudi07}. Another one involves a
second-order phase transition below a certain critical temperature
\cite{Salam91} that could work at low temperatures such as those
of the interstellar space. However, theoretical or experimental
conclusive results from the diverse mechanisms to amplify
enantioselection based on electroweak energy difference, are not
yet at hand (e.g.,\cite{Cintas00,Chandra08}).

 Another universal mechanism that could
discriminate between the two enantiomers of a chiral molecule and
that involves weak interaction is neutrino-electron axial-vector
interaction mediated by charged electroweak bosons W$^{\pm}$. This
process discriminates on the basis of an asymmetry between the
number of neutrinos and antineutrinos, and that electrons of
opposite (L,R) enantiomers have opposite helicity. Considering the
cosmological relic sea of neutrinos, the estimated energy
difference between the two enantiomers was found to be even lower
than the value $10^{-21}$ eV mentioned above \cite{Bargue06}.
However, it could increase significantly under bigger neutrino
fluxes, as in supernova remnants \cite{Bargue07} assumed that big
molecules could survive in the surroundings. Revisited assumptions
about the number density of cosmological neutrinos in the relic
sea lead to an increase of the mentioned energy difference, as we
shall see in the next section.

Looking for other universal mechanisms acting also outside of the
Earth, we analyze here the possible enantioselective effect of
chiral dark matter on chiral molecules.

The existence of dark matter is inferred from astrophysical
observations in light of studies of the dynamics of stars in the
local disk environment, rotation curves for a large number of
spiral galaxies, gravitational lensing by clusters of galaxies and
some large scale studies of the Universe (for a recent review of
experimental searches for dark matter see for example
\cite{Munoz04}). A vast variety of candidates have been proposed
for dark matter content, from baryonic to non-baryonic matter. The
non-baryonic candidates are basically postulated elementary
particles beyond the Standard Model which have  not been
discovered yet, like axions, WIMPs (Weak Interacting Massive
Particles) and other exotic candidates.
 The baryonic candidates are the Massive Compact Halo Objects (Macho)
\cite{Sadoulet99}. Another important difference is the hot versus
cold dark matter. A dark matter candidate is called hot if it was
moving at relativistic speeds at the time when galaxies could just
start to form, and cold if it was moving non-relativistically at
that time. The problem is that hot dark matter cannot reproduce
correctly the observed structure of the Universe. Therefore we
focus our attention on cold dark matter. The fact that dark matter
interacts weakly with matter makes its detection very difficult
\cite{Munoz04}. However many experiments are currently in progress
in order to reach this goal.

Here we estimate the energy difference between the two enantiomers
of a chiral molecule, due to the weak type interaction between
non-baryonic cold dark matter (specifically WIMPs)  and molecular
electrons with non zero helicity. Experimental results on dark
matter are used. Given the resemblance in the procedure with the
energy difference induced by  cosmological neutrinos, estimated in
a previous work \cite{Bargue06}, we  first recall this procedure
at the time we improve the result we obtained in that work.

\section{Energy difference between opposite enantiomers induced by cosmological neutrinos}

Following a previous work \cite{Bargue06}, we consider
neutrino-electron interactions mediated by the axial-vector
Hamiltonian density
     \begin{equation}
      H=\frac{G_{F}}{\sqrt{2}}\bar e \gamma^{\mu}(g_{V}-g_{A}\gamma_{5})e\bar \nu
       \gamma_{\mu}(1-\gamma_{5})\nu,
       \label{eq2}
     \end{equation}
\noindent where $G_{F}$ is Fermi's constant,  $e(\nu)$ denotes the
electron (neutrino) spinor field, $\bar e (\bar\nu)$ is its
adjoint spinor, $\gamma^{\mu}$ are the Dirac matrices (regarded as
a four-vector),
$\gamma^{5}=i\gamma^{0}\gamma^{1}\gamma^{2}\gamma^{3}$, and
$g_{V,A}$ are suitable coupling constants that parameterize the
strength of the interactions. As it was discussed for example in
\cite{Gelmini05}, in the non-relativistic limit it is possible to
 make the following approximations for the dominant temporal
 components of the four-vectors appearing in the above
 Hamiltonian,
     \begin{eqnarray}
\bar e \gamma^{\mu} \gamma_{5} e &\sim& \vec{\sigma}_{e} \cdot
\vec{v}_{e} , \qquad \
\nonumber\\
\bar \nu \gamma_{\mu} \nu &\sim& n_{\nu}-n_{\bar\nu} \qquad
(\textrm{Dirac neutrinos}),
\nonumber\\
\bar \nu \gamma_{\mu}\gamma_{5} \nu &\sim& n_{\nu_l}-n_{\nu_r}
\qquad (\textrm{Majorana neutrinos}), \label{eq3}
      \end{eqnarray}
where the number density differences, $n_{\nu}-n_{\bar \nu}$ and
$n_{\nu_l}-n_{\nu_r}$ refer to neutrino-antineutrino and
left-right helicity eigenstates respectively. Obviously they are
not zero only in the case where  there is a net lepton number or
helicity in the cosmic neutrinos background. We recall from
\cite{Bargue06} and \cite{Stod75} that for Dirac neutrinos the
energy splitting obtained for the electron is
     \begin{equation} \Delta E \sim
G_{F} |(n_{\nu}-n_{\bar\nu}) \langle\vec{\sigma}_{e} \cdot
\vec{v}_{e}\rangle|, \label{eq4}
     \end{equation}
where the expected value of the electron helicity $\langle
\vec{\sigma}_{e} \cdot \vec{v}_{e}\rangle$ takes opposite signs
for the two opposite enantiomers, as we can see from a simplified
chiral molecule model \cite{Perez91}. In this model, a dominant
axial symmetry around axis Z, with a left(right)-handed
perturbative potential of period a, is assumed, so that the
electronic molecular states can be described by superposition of
eigenstates of both angular momentum $L_z$  and linear momentum
$P_z$ (eigenvalues $\hbar n(2 \pi/a)$), i.e. $|M_L,n\rangle$, in
the form
               \begin{eqnarray}
            \Phi_L &=&c_0 |0,0\rangle
                         +c_1 |+1,-1\rangle + c_2
                         |-1,+1\rangle
                         \, \label{FUNL},\\
             \Phi_R &=&c_0 |0,0\rangle
                         +c_1 |+1,+1\rangle + c_2 |-1,-1\rangle
                         \, , \label{FUNR}
            \end{eqnarray}
with $|c_1|^2=|c_2|^2 \equiv  C $. These L and R states have then
opposite helicities:
            \begin{equation}
             \langle \Phi_L |  L_z P_z | \Phi_L \rangle= -4 C \pi
             /a =  -  \langle \Phi_R |  L_z P_z | \Phi_R \rangle\, . \label{CHIR}
            \end{equation}
Notice that we are using all the time natural units where
$\hbar=c=1$. The spin of the electron can be taken into
consideration by replacing $L_z$ with $J_z=L_z + S_z$. In a
realistic chiral molecule the electronic states would not be
eigenstates of the helicity, but its mean value would have
opposite sign for L and R enantiomers. The parameter $C<1/2$
accounts for the degree of chirality.

We note that the velocity of the molecule carrier (interstellar
grains, meteorites, the Earth...) does not contribute to the
helicity of the electrons: If $\vec P_T$ is the translational
momentum of the carrier, the electronic wave function $
\Phi_{L(R)}$ must include the factor $e^{i\vec P_T \cdot \vec R}$
(here $\vec R$ is the  position of the molecule), and the
contribution of $\vec P_T$ to the electron helicity is then
{\setlength\arraycolsep{1pt}
\begin{eqnarray}
&&\langle \Phi_{L(R)}e^{i\vec P_T \cdot \vec R} |\vec L \cdot \vec
P_T| \Phi_{L(R)}e^{i\vec P_T \cdot \vec R} \rangle
\nonumber\\
&& =\langle \Phi_{L(R)}|\vec L|\Phi_{L(R)}\rangle \cdot \vec P_T =
0 \, ,
\end{eqnarray}}
since $\langle \Phi_{L}|\vec L|\Phi_{L}\rangle =\langle
\Phi_{R}|\vec L|\Phi_{R}\rangle=0$ as can be seen from Eqs.
(\ref{FUNL},\ref{FUNR}). We also remark that the particle flux is
assumed to be isotropic, thus, its interaction with the electrons
of a chiral molecule is the same irrespective of the orientation
of the molecule.

The energy difference that we obtained, assuming complete
neutrino-antineutrino asymmetry, with number density  of about
$10^{-2} \mathrm{cm}^{-3}$, $C=1/2$, $a \sim 1$ Angstrom and the
electron helicity given by Eq. (\ref{CHIR}), was of the order of
$10^{-26}$ eV \cite{Bargue06}.

However, it has been recently suggested \cite{Gelmini05,Duda01}
that, in scenarios beyond the standard model, the
neutrino-antineutrino density asymmetry $n_{\nu}-n_{\bar \nu}$
could be up to the order of $\sim 10 - 1050 \ \mathrm{cm}^{-3}$.
Although the extreme upper bound density asymmetry seems to be
excluded by considerations of primordial nucleosynthesis
\cite{Dolgov2002}, we consider it to estimate an upper bound of
the energy difference.

If we take the value $n_{\nu}-n_{\bar \nu} \sim 1000 \
\mathrm{cm}^{-3}$, we then obtain an
upper bound for the energy difference between enantiomers of the
order of $10^{-21}$ eV, per molecular electron with non zero
helicity.
Evidently this tiny energy needs massive amplification mechanisms
as those mentioned at the beginning in order to induce an
effective enantioselection.

\section{Energy difference between opposite enantiomers induced by
fermionic cold dark matter}

In a similar way to the neutrino-electron interaction above
considered, we are now to estimate the electron energy splitting
induced by the axial-vector interaction between a fermionic dark
 matter candidate (typically a WIMP) and an electron. The relevant Hamiltonian
  density can be
 written  as
    \begin{equation}
H =   \sum_i  d_i  \bar \chi \gamma_{\mu}(1-\gamma_{5}) \chi \bar
\psi_i \gamma^{\mu}\gamma_{5}\psi_i,
     \end{equation}
where $\chi$ is the dark matter spinor which can be Dirac or
Majorana. The index $i$ runs through $i=e,u,d,s$, i.e., we are
considering also the interaction between the dark matter particle
and the $u,d$ and $s$ quarks. This will be important later in
order to use the present experiments trying to measure the WIMP
flux on Earth to set some bounds on the possible effect of  dark
matter on opposite enantiomers. Therefore $d_e, d_u, d_d$ and
$d_s$ are the coupling of the $\chi$ field to the different matter
fields $e=\psi_e, u=\psi_u, d=\psi_d$ and $s=\psi_s$.

 As  WIMPs  are typical examples of cold dark matter and heavy by
 definition, we can invoke again the non-relativistic limit. Thus, as
 it was the case of neutrinos, for  Dirac   WIMPs the term
$ \bar \chi \gamma_{\mu} \chi$ dominates with the temporal
component of this vector being proportional to $n_\chi-n_{\bar
\chi}$. For the Majorana case only the axial vector $ \bar \chi
\gamma_{\mu} \gamma_{5} \chi$ remains and its temporal component
becomes proportional to $n_{\chi_l}-n_{\chi_r}$, as in Eq.
(\ref{eq3}).

The expression for the corresponding electron energy splitting is
similar to that of Eq. (\ref{eq4}),
    \begin{equation}
\Delta E \sim  d_e |\Delta n \langle\vec{\sigma}_{e} \cdot
\vec{v}_{e}\rangle|, \label{eq12}
     \end{equation}
where $\Delta n$ is the appropriate number density difference
corresponding to the Dirac or the Majorana case. In principle
these differences depend on the unknown nature of the dark matter
and its evolution along the universe history. In the following we
will write these differences as $|\Delta n|= \alpha n$ where $n$
is the total WIMP number density. Clearly the parameter $\alpha$
is a measure of the degree of particle-antiparticle or left-right
asymmetry present in the the dark matter respectively. For
example, in the case of Dirac dark matter, $\alpha=1$ indicates
that all WIMPs are particles with no antiparticles present and
$\alpha=0$ means a complete particle-antiparticle symmetry. As in
the neutrino case, interactions between molecular electrons with
non zero helicity and cold dark matter could lead to an energy
difference between the two enantiomers of a chiral molecule
whenever the parameter $\alpha$ is different from zero. To have an
estimation of the energy difference, we consider the interaction
between WIMPs and an electron of a chiral molecule. Let $\rho = n
M_\chi$ be the energy density of those WIMPs, with $M_\chi$ being
their mass and $n$ their number density. The density of WIMPs
trapped in the gravitational potential wall of the galaxy is
expected to be of the order of $\rho \sim$ 0.3 GeV
$\mathrm{cm}^{-3}$. Then the energy splitting can be written as
  \begin{equation}
\Delta E \sim   d_e\alpha \
\frac{\rho}{M_\chi}|\langle\vec{\sigma}_{e} \cdot
\vec{v}_{e}\rangle| \, .
  \end{equation}
In order to see how important this splitting could be, we need to
know which values of the coupling constant $d_e$  are acceptable.
In principle there is no any available experimental information
about $d_e$. However one reasonable assumption that could be done
is that all the $d_i$ couplings are at least of the same order of
magnitude. In the absence of a theory of WIMPs this seems to be
not so bad assumption since WIMPs does not interact strongly with
matter. If this is the case, one can then use the present bounds
on the elastic cross-section proton$-\chi$ to get some information
about the size of the $d_i$ couplings. In order to compute this
cross section, one needs to relate the quark$-\chi$ couplings with
the proton$-\chi$ coupling. This can be done by using the
effective Hamiltonian (see \cite{Binetruy} and references therein)
   \begin{equation}
H = -a_p2\sqrt{2} \bar \chi \gamma_{\mu}\gamma_{5} \chi \bar p
s^\mu p \, ,
\end{equation}
where $p$ is the proton spinor and  $s^\mu$ is its spin vector
(here we are considering the Majorana case but the Dirac case can
be treated in a similar way). The coupling $a_p$ is defined as
\begin{equation}
a_p=\frac{1}{\sqrt{2}}\sum_{i=u,d,s} d_i \Delta q_i^{(p)} \, .
\end{equation}
The constants $\Delta q_i^{(p)}$ (with $q_1=u, q_2=d$ and $q_3=s$)
are introduced trough the proton matrix element
\begin{equation}
\langle p|\bar \psi_i \gamma^\mu\gamma_5 \psi_i  |p\rangle  = 2
s^\mu \Delta q^{(p)}.
\end{equation}
Experimentally we have $\Delta u^{(p)}\simeq0.78$, $\Delta
d^{(p)}\simeq -0.5$ and $\Delta s^{(p)}\simeq -0.16$. Then, by
using standard methods, it is straightforward to compute the
elastic proton$-\chi$ cross-section, which is given in the proton
rest frame by
\begin{equation}
\frac{d \sigma}{d q^2} = \frac{\sigma_n}{4 v_\chi \mu^ 2} \, ,
\end{equation}
where $\vec q$ is the momentum transfer, $v_\chi$ is the $\chi$
velocity, $\mu$ is the proton $-\chi$ reduced mass and
\begin{equation}
\sigma_n = \frac{12 a_p^2m_p^2M_\chi^2}{\pi(m_p+M_\chi)^2}
\end{equation}
($m_p$ being the proton mass) is just the non-relativistic cross
section for vanishing momentum transfer. Now days there are many
experiments around the world trying to detect WIMPs directly
(visit the webpage \cite{dmtools} for complete and upgraded report
of their main results). Usually they set exclusion regions on the
plane $\sigma_n-M_\chi$. From the recent XENON10 2007
\cite{dmtools} we learn for example that, for $M_\chi\simeq 100$
GeV, $\sigma_n$ must be lesser than $10^{-43}$ cm$^2$ and, for
$M_\chi\simeq 1000$ GeV, lesser than $10^{-42}$ cm$^2$. Assuming
for simplicity all the quark couplings to be the same, i.e.
$d_q\simeq d_u\simeq d_d \simeq d_s$, we have $a_p \simeq 0.0072
d_q^2$. Then we get that for $M_\chi\simeq 100$ GeV, $d_q^2 <
10^{-14}$ GeV$^{-4}$ and for $M_\chi\simeq 1000$ GeV, $d_q^2 <
10^{-13}$ GeV$^{-4}$. As discussed above we now assume $d_e \sim
d_q$. Then it is possible to set a bound on the energy splitting
which turn to be very tiny even in best case corresponding to
$M_\chi\simeq 100$ GeV. We obtain in this case, with an electron
velocity about $10^{-2}$,
 $\Delta E \leq \alpha 10^{-44}$ eV.

\section{Conclusion}

We have analyzed the effect of cold dark matter on the
discrimination between the two enantiomers of a chiral molecule
whose  external electrons have opposite helicities in the
respective opposite enantiomers. The estimated energy difference
between the two enantiomers, due to WIMP-electron interaction, is
found to be extremely small, several orders of magnitude lower
than that induced by electron-nuclei weak interaction.  Hence,
dark matter would  be discarded as inductor of chiroselection
between enantiomers and then of biological homochirality. By
contrast, the enantioselective  effect of the cosmological relic
sea of neutrinos acquires relevance with the current assumptions
about the number density of cosmological neutrinos. In this case
we obtain an energy difference
between $10^{-23}$ and $10^{-21}$ eV for  the two opposite
enantiomers
per molecular electron with non zero helicity.
The upper bound of the energy difference, although could be
excluded by reasons previously mentioned, reaches the same order
of magnitude as the energy difference induced by the well known
electron-nucleus electroweak interaction in some molecules.

{\large acknowledgments}

This work is supported by DGICYT (Spain) project
 BPA2005-02327, by the Universidad Complutense/CAM projects
910309 and CCG06-UCM/ESP-137, and by the MEC (Spain) projects
CTQ2005-09185-C02-02 and FIS2004-03267. The work of P. Bargue\~no
was supported by the FPI grant BES-2006-11976 from the Spanish
MEC. The authors would like to thank A. L. Maroto and R. P\'erez
de Tudela for useful discussions.

\end{document}